\documentclass[
    reprint,
    superscriptaddress,
    amsmath,
    amssymb,
    aps,
    prl,
    notitlepage
]{revtex4-2}

\usepackage{physics}
\usepackage{enumitem}

\usepackage{graphicx}
\usepackage{dcolumn}
\usepackage{bm}
\usepackage{dsfont}
\usepackage[colorlinks=true,urlcolor=blue,linkcolor=blue,citecolor=magenta]{hyperref}
\usepackage[normalem]{ulem}
\usepackage{tikz}
\usetikzlibrary{decorations.pathreplacing, arrows.meta, calc}
\usepackage{amsmath, amssymb, amsthm, mathtools, stackrel}
\usepackage[T1]{fontenc}
\usepackage{bbold}
\usepackage{standalone}

\usepackage{algorithm2e}

\newcommand{\rz}{\right]}
\newcommand{\lz}{\left[}
\newcommand{\rd}{\right\}}
\newcommand{\ld}{\left\{}
\def\({\left(}
\def\){\right)}

\newcommand{\avg}[1]{\langle#1\rangle}

\newcommand{\customref}[2]{\hyperref[#1]{\ref*{#1}#2}}

\definecolor{Ured}{HTML}{cc0000}
\definecolor{Ublue}{HTML}{1f65cf}
\definecolor{Ugreen}{HTML}{198a11}

\newcommand{\BB}{\mathcal{B}}
\newcommand{\CC}{\mathcal{C}}

\theoremstyle{definition}

\usepackage{nicefrac}

\usepackage[capitalize,nameinlink]{cleveref}

\graphicspath{ {img/} }

\makeatletter
\def\l@subsubsection#1#2{}
\makeatother

\begin{document}

\title{Kinetics of sliding-window quantum error correction}

\author{Adithya Sriram}
\affiliation{Department of Physics, Stanford University, Stanford, CA 94305}
\author{Charles Stahl}
\affiliation{Department of Physics, Stanford University, Stanford, CA 94305}
\author{Aleksander Kubica}
\affiliation{Yale Quantum Institute \& Department of Applied Physics, Yale University, New Haven, CT 06520}
\author{Yaodong Li}
\affiliation{Department of Physics, Stanford University, Stanford, CA 94305}
\affiliation{Department of Mechanical Engineering \& Department of Physics, University of Washington, Seattle, WA 98195}
\date{August 11, 2026}

\begin{abstract}

Practical implementations of quantum error correction (QEC) require rapid measurement and continuous processing of 
the syndrome information 
in order to prevent a backlog of unprocessed data.
While ``static'' QEC is theoretically well understood via mappings to equilibrium statistical mechanics models, such an understanding of ``real-time'' QEC is currently lacking.
Here, we study the kinetics of
sliding window decoding (SWD), an implementation of real-time decoding that acts on temporally local windows of noisy syndrome information and commits to corrections irreversibly at a nonzero rate.
We propose an effective description of SWD in terms of a stochastic kinetic process, where $\mathbb{Z}_2$-charged point particles undergo parity-conserving reaction and diffusion.
This model describes dynamics at length and time scales large compared to the window size $W$,
whereas physics at scales smaller than $W$ leads to nontrivial $W$-dependent scaling of the effective parameters.
We identify the rate of decoding $1/W$ as a relevant perturbation to the decodable phase.
We also show broad applicability of our results by changing many microscopic details of SWD without affecting the effective description.
\end{abstract}

\maketitle

\begin{figure*}[t]
    \centering
    \includegraphics[width=\linewidth]{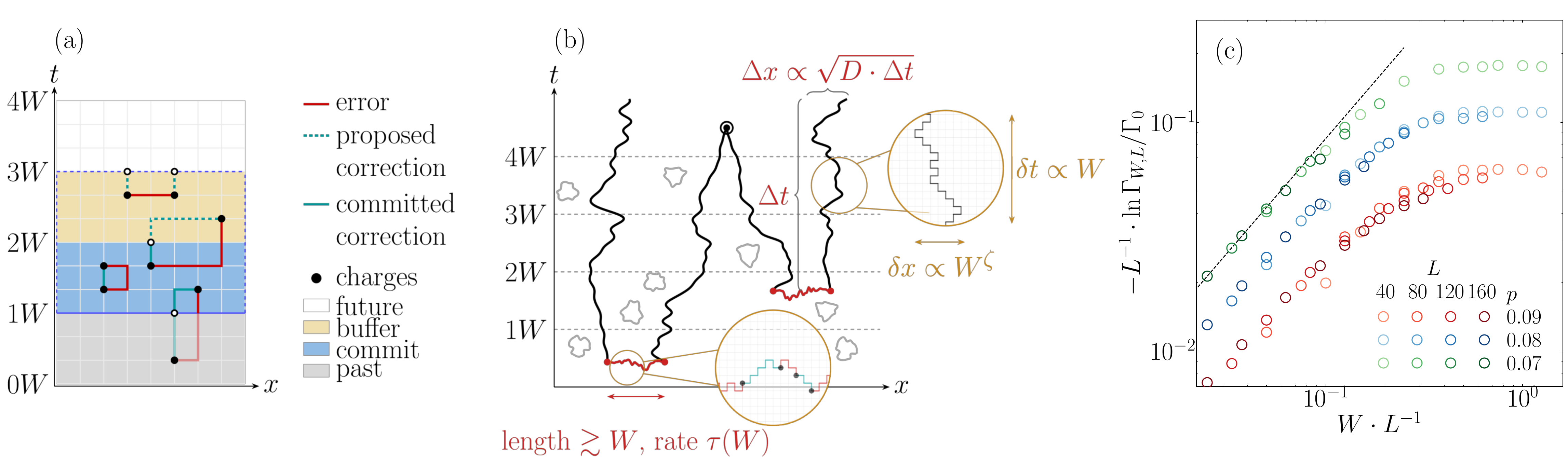}
    \caption{
    (a) Illustration of the sliding-window decoder, for the first few rounds.
    The error and the correction together form closed loops in the bulk, with possible termination on the temporal boundary between the ``buffer'' and the ``future'' regions.
    Within each round, only correction operations in the ``commit'' region are turned into physical corrections; those in the ``buffer'' region serve as a look-ahead to improve decoding accuracy.
    (b)
    Worldlines of $\mathbb{Z}_2$ charges.
    \textit{Fast} pair nucleation and annihilation events occur within $O(W)$ time scale (gray), whereas \textit{slow} processes persist over many decoding rounds (black), and contribute to decoding failure.
    Nucleation events for slow charges are highlighted in red.
    (c) Data collapse for the memory time extracted from $\mathbb{P}_{\rm fail}(t; W, L)$ against~\cref{eq:crossover_scaling}, see main text for details. Results at different values of $p < p_c$ fall onto the same universal function $\Phi$ (up to an overall rescaling of the vertical axis).
    }
    \label{fig:placeholder_fig_1}
\end{figure*}

\emph{Introduction} --- There has been tremendous progress in bringing quantum error correction (QEC) from a question of theoretical importance to practical 
implementation~\cite{RyanAnderson2024,Acharya2024,Pogorelov2025,Putterman2025,Lacroix2025,SalesRodriguez2025,Brock2025,Daguerre2025,Eickbusch2025,Bluvstein2025,Zhang2026}.
Protecting quantum information in complex logical circuits while maintaining fast logical clock speeds is vital as we transition into the fault-tolerant era.
To analyze the properties of a code, one often imagines an idealized ``static'' setting in which the experimentalist records (possibly faulty) syndrome information for an extensive period of time; this full history is used to infer the location of errors and make decoding decisions. 
Most of the theoretical understanding of QEC is based on this idealized setting.
For instance, the seminal work of Dennis et al.~\cite{DKLP2001topologicalQmemory} relates such a problem to a disordered equilibrium spin model and its phase transitions, with general applicability to a large family of QEC codes~\cite{katzgraber2009,bombin2012,pryadko2014, kubica-2018-PRL-statmech-3Dcolorcode,ChubbFlammia2018, li2024perturbative}.

In contrast, any practical QEC protocol will require some version of ``real-time'' decoding, in which one only has access to the syndrome information within a short time interval before a decoding decision must be made. 
For example, in a scenario in which logical operations must be performed on the quantum information at a nonzero rate, 
decoding speed becomes the bottleneck~\cite{terhal2015RMP}. %
In a prototypical
scheme of real-time QEC, known as sliding-window decoding (SWD)~\cite{DKLP2001topologicalQmemory, skoric2023parallel, tan2022parallel, Battistel2023review}, the decoder combines a new set of syndrome information with the previously stored one, commits to corrections, and then discards the old data.
This way, the decoder only ever maintains a finite amount of syndrome information. %
Practical considerations behind real-time decoding schemes have motivated many technical improvements~\cite{skoric2023parallel, tan2022parallel, bombin2023modular},
but comparatively little attention has been devoted to developing their theoretical descriptions. 

Two features of real-time QEC rule out the equilibrium statistical mechanics description of the static setting.
First, decoding decisions are sequential and irreversible, made at a finite rate. Second, 
errors compound over time through a nontrivial interplay of physical noise, measurement noise, and imperfect recovery, so that the accumulated error is not captured by any local noise model.
Consequently, one may wonder whether real-time QEC can be related to any effective model, and whether such a model must be intrinsically non-equilibrium.
Such a connection would not only provide a physical intuition and effective description of the dynamics of QEC, but also establish fundamental bounds on the performance of real-time QEC, providing a theoretical benchmark for recent numerical results
~\cite{skoric2023parallel, tan2022parallel, bombin2023modular, oberoi2026adaptadaptivewindowdecodingpractical, huang2023increasing}.

In this work, we develop such a description for SWD %
for topological codes with deconfined point-like defects carrying $\mathbb{Z}_2$ charge.
Through analytical arguments and numerical experiments, we show that SWD, at time and length scales larger than the window size $W$, admits an effective description in terms of a stochastic %
kinetic model, namely a parity-conserving reaction-diffusion process, see~\cref{eq:RD}.
The reaction-diffusion kinetics might be anticipated on symmetry grounds, and we show in the End Matter that indeed many microscopic details of SWD can be changed without modifying this universality class.
On the other hand, the scaling relations between parameters of the reaction-diffusion process and $W$ are nontrivial and are not determined by symmetry alone.
We connect the microscopic SWD dynamics and the effective description by extracting such scaling relations, and by describing a crossover between the real-time regime (where $W = O(1)$) and the static regime (where $W$ exceeds the code distance) with $1/W$ as a relevant parameter.
Our work thus yields a greater theoretical understanding of the intrinsic tradeoffs one must grapple with in real-time decoding.

\emph{Definition of SWD} --- We first describe SWD in detail for a general $d$-dimensional CSS code of linear size $L$ under phenomenological noise with physical and measurement errors of rate $p$.
We will then focus on the 1D repetition code and the 2D toric code as prototypical examples in our numerical studies.
For simplicity, we assume the decoder used is minimum weight perfect matching (MWPM)~\cite{DKLP2001topologicalQmemory}, though we also explore other decoder choices in the End Matter.
Errors in the code create $\mathbb{Z}_2$ point-like charges and we will refer to them as such. 
In the Discussion, we will also comment on SWD as applied to more general classes of codes.

As we illustrate in~\cref{fig:placeholder_fig_1}(a),
SWD proceeds via consecutive rounds of decoding, which we label with the integer $n \in [0,N]$.
During the $n$-th round, we consider the syndrome information within the time window $[nW, (n+2)W]$,
which reveals the locations of $\mathbb{Z}_2$ charges.
The syndrome is input to the decoder, which matches the measured error chain endpoints, i.e. charges. Now, we divide this spacetime slice into a \textit{commit window} and a \textit{buffer window}, each of temporal size $W$~\footnote{In principle, the relative size of the buffer window and commit window may be tuned, but for simplicity, we make them equal.}.
The decoder only applies corrections that lie in the commit window.
If the decoder matches charges that span the temporal boundary between the commit and buffer windows, a truncation of the matching to the commit window leaves behind a charge on this temporal boundary [hollowed dot in~\cref{fig:placeholder_fig_1}(a)], and this charge persists into the next round.
The buffer window serves as a look-ahead that improves decoding accuracy, but no final decisions are made there yet.
After this round, syndrome information in the commit window is erased.
Each round thus overlaps with its neighboring ones and corrections are made every $W$ time steps at a rate $1/W$.
This process is depicted for the (1+1)D repetition code in \cref{fig:placeholder_fig_1}, and a formal definition can be found in the Supplemental Material~\footnote{See Supplemental Material for further discussions of (i) formal definition and effective description of the sliding window decoder (SWD), and (ii) an absorbing state phase transition of the SWD under SWAP noises.}.

The combination of the errors and the proposed and committed corrections form closed loops in spacetime, except at the temporal boundary between the buffer window and the future window where open loops can anchor.
These loops can be interpreted as the worldlines of the charges.

\emph{Effective model for SWD} ---
In order to arrive at an effective description of the dynamics, we must make a crucial distinction between the \textit{fast} $\mathbb{Z}_2$ charges which are created and annihilated within a single commit window, 
and the \textit{slow} $\mathbb{Z}_2$ charges which actually contribute to logical failure. 
The worldlines of fast charges form closed loops of small spatiotemporal extent (see \cref{fig:placeholder_fig_1}(a)) and they contribute a nonzero background density at any instant. However, they do not grow or persist and therefore do not seed logical failure.

On the other hand, the slow charges which do cause logical failure are those whose initial spatial distance reaches $\geq W$ within a single commit window. For a pair of such charges, the action of the decoder on these charges, if there does not exist a suitable matching involving other neighboring charges, transports each of them 
a bounded distance (of order $W^\zeta$ per round, see~\cref{eq:diffusion_constant_result})
independent of $L$
(\cref{fig:placeholder_fig_1}(a)).
The pair therefore can persist over many rounds of decoding, essentially performing independent random walks with finite step sizes. The time at which these pairs traverse the system and cause a logical error is what we call the memory time $t_{\rm mem}$.
We provide more details on their quasi-local motion in the Supplemental Material~\cite{Note2}.

We argue that the motion of these slow charges is described by the following reaction-diffusion process~\cite{CardyTauber1998, henkel2008non, Krapivsky_Redner_Ben-Naim_2010}
\begin{subequations}
\label{eq:RD}
\begin{align}
\label{eq:reaction}
   & \text{reaction: }\quad A+A \xrightleftharpoons[\tau]{\lambda} \emptyset,\\
\label{eq:diffusion}
   & \text{diffusion: }\quad \emptyset + A \xleftrightarrow{D} A +\emptyset, 
\end{align}
\end{subequations}
where $A$ refers to a $\mathbb{Z}_2$ charge and $\emptyset$ refers to vacuum. \cref{eq:reaction} describes a process in which two charges annihilate at rate $\lambda$ to produce a vacuum state, as well as the reverse, in which two charges nucleate from a vacuum state at rate $\tau$. 
\cref{eq:diffusion} describes the diffusion process, in which a charge can move to neighboring unoccupied sites at rate $D$.
While pair creation is predominantly a consequence of the errors, diffusion and pair annihilation come from both the errors and the decoder.

We estimate the pair nucleation rate of the slow $\mathbb{Z}_2$  charges via a standard Peierls argument for MWPM, provided a sufficiently small $p$. 
In order for a pair to be created and persist beyond a single commit window, $O(W)$ ``error events'' are required to create and diffuse the charges a distance $\geq O(W)$ apart from each other. 
This leads to a pair nucleation rate $\tau \propto p^{\kappa W}$ at leading order, 
where $\kappa$ is a nonuniversal geometrical factor that is fixed by the decoder.
Generalizing to include higher order effects, we obtain the pair nucleation rate
\begin{align} \label{eq:tau_W_scaling}
    \tau(W) \propto e^{- \kappa \sigma W},
\end{align}
where $\sigma = \sigma(p)$ is the line tension that depends only on the error rate $p$ but not $W$ or $L$.

When two charges meet, they annihilate and cease to extend in the temporal direction.
We therefore take the annihilation rate $\lambda(W) = 1$ independently of $W$, describing annihilation upon contact with probability $1$.

We can also extract a diffusion constant by treating the motion of the coarse-grained charges as a random walk whose elementary moves are the worldlines within each window.
These worldlines traverse a strip of temporal extent $W$. The microscopic processes within a window cause the worldline to acquire some spatial displacement $\delta \mathbf{r}$, whose statistics follow that of directed polymers in random media (DPRM), $|\delta \mathbf{r} |  \propto W^\zeta$,
where $\zeta$ is the wandering exponent~\footnote{In (1+1)D and for a large enough $W$, the worldlines here are domain walls within a 2D random-bond Ising model, and they have DPRM exponents~\cite{huse_henley_1985_DPRM, kardar1985roughening, huse_henley_fisher_1985_respond, KPZ1986}.}.
The coarse-grained charge therefore executes a random walk with step size $|\delta \mathbf{r} | \propto W^\zeta$ taken every $\delta t \propto W$ time units, see \cref{fig:placeholder_fig_1}(b).
Within $\Delta t$ time, the large-scale random walk takes $N = \Delta t / \delta t$ such steps, and travels a distance $|\Delta \mathbf{r}| \propto |\delta \mathbf{r}| \cdot N^{1/z}$,
where $z = 2$ is the dynamic exponent of diffusion.
We thus have an effective diffusion constant %
\begin{align} \label{eq:diffusion_constant_result}
    D(W) \propto | \delta \mathbf{r}|^z \cdot (\delta t)^{-1} \propto W^{z\zeta - 1}.
\end{align}
For SWD on the repetition code in (1+1)D we have the DPRM exponent $\zeta = 2/3$~\cite{huse_henley_1985_DPRM, kardar1985roughening, huse_henley_fisher_1985_respond, KPZ1986}.

Putting these processes together, we have a simple effective description of the dynamics in terms of a reaction-diffusion process of $\mathbb{Z}_2$ charges. The parameters $D$ and $\tau$ of the effective model scale with the window size $W$ nontrivially as described in \cref{eq:diffusion_constant_result} and \cref{eq:tau_W_scaling}. We expect these to hold whenever $p < p_c$~\footnote{Here, $p_c$ is the threshold
defined in the static limit ($W \to \infty$), 
as our arguments really only depend on the assumption of a nonzero domain wall line tension. %
}.
We also expect that our discussions apply generally to codes with deconfined point-like excitations in an arbitrary spatial dimension.

The scaling of $\tau$ has direct consequences on the memory time of the code.
With $\tau > 0$, charges nucleate and proliferate.
At times longer than $t_{\rm relax}$, the reaction-diffusion process reaches a steady state with a nonzero charge density.
Similarly, at times longer than $t_{\rm mem}$, the code enters a state where different logical sectors are maximally mixed.
For the codes that we consider here, a finite density of charges leads to a nonvanishing logical error rate, so we intuitively identify the relaxation time $t_{\rm relax}$ of the process in \cref{eq:RD} with the memory time $t_{\rm mem}$ of the code under the SWD dynamics~\footnote{However, in single-shot codes, self-correcting codes, or other code families this identification may not hold.}.
In particular, a numerical calculation of the memory time also provides an indirect test of \cref{eq:tau_W_scaling} via 
\begin{align}
\label{eq:tmem_tau_from_RD_dynamics}
    t_{\rm mem} \propto t_{\rm relax} \propto \tau^{-\alpha} \propto e^{+\alpha \kappa \sigma W},
\end{align}
where $\alpha$ is a universal exponent for the process in~\cref{eq:RD}~\cite{racz1985}, and we have $\alpha = 1$ in one spatial dimension.
Below, we numerically test the exponential scaling of $t_{\rm mem}$ with $W$ without directly extracting $\alpha$~\footnote{In our numerical experiments, slow charges are always clouded by a background of fast charges, and are not easily distinguishable.
}.

\emph{Crossover to $W \gg L$} --- 
Throughout the above discussion, we have assumed that $W \ll L$.
If instead $W\to \infty$ we see from \cref{eq:diffusion_constant_result} that $D$ diverges, indicating the breakdown of the kinetic description. 
The charges are no longer random walkers slowly diffusing across the system, as they can now make nonlocal jumps.
Indeed, as $W$ approaches and exceeds $O(L)$,
the reaction-diffusion dynamics~\cref{eq:RD} must give way to the equilibrium statistical mechanics model.
Here, the dominant logical error is no longer due to proliferation of $\mathbb{Z}_2$ charges, but instead a single space-like domain wall crossing the system, whose free energy cost scales linearly in $L$. This yields a memory time that diverges as $ \ln t_{\rm mem} \propto L$.

Our discussions above motivate the following scaling form describing the crossover in $W$,
\begin{align} \label{eq:crossover_scaling}
    \ln t_{\rm mem} \propto L \cdot \Phi(W/L),
\end{align}
for a \textit{universal} scaling function $\Phi$. 
To match the scaling of $t_{\rm mem}$ for $W \ll L$ and $W \gg L$,  the limiting behaviors of $\Phi(x)$ are required to be (1) for $x \to 0$, $\Phi(x) \approx \alpha \kappa \sigma x$, so that $\ln t_{\rm mem} \approx \alpha \kappa \sigma W$ in agreement with~\cref{eq:tmem_tau_from_RD_dynamics}, and (2) for $x \to \infty$, $\Phi(x) \to \sigma$, so that $\ln t_{\rm mem} \approx \sigma L$, in agreement with the expectation in the static limit.

\emph{Numerical results} --- We now provide numerical tests of SWD for the 1D repetition code.
These results also provide evidence for the effective kinetic description from above.
Parameters of this problem include the system size $L$, the window size $W$, the error rate $p$, and
the time steps $t$, which is taken to be up to multiples of $W$. We will fix $p$ to be below the threshold $p_c$ (defined when $W \to \infty)$ and vary the other parameters. 
For the repetition code, we have $p_c \approx 0.103$.
We also study the 2D toric code, with results reported in End Matter.

After $t$ time steps (i.e. $t/W$ rounds of SWD), we define the logical failure probability at this time, denoted $\mathbb{P}_{\rm fail}(t; W, L)$,
through a noiseless recovery channel (chosen to be MWPM without measurement errors) applied to the final state.  
For the repetition code, it is also useful to define $\Delta \rho \coloneqq |\rho_0 - \rho_1|$, i.e. the absolute value of the 
\textit{spin density}
in the physical state. 
We initialize the system in the all-0 state, i.e. $\Delta \rho = 1$. We track both $\Delta \rho (t; W,L)$ and $\mathbb{P}_{\rm fail}(t; W,L)$ over time.
Our choice of these observables, which track the physical and logical errors in the system, is motivated by previous studies of single-shot QEC~\cite{bombin2015singleshot, KubicaVasmer, BrownNickersonBrowne2016GCC,Gu2024}.

The behavior of $\mathbb{P}_{\rm fail}(t; W,L)$ is characterized by a linear-in-$t$ growth with slope denoted $\Gamma_{W,L}$, up until some time scale $\propto \Gamma_{W,L}^{-1}$, at which point $\mathbb{P}_{\rm fail}(t; W,L)$ saturates~\footnote{The steady state value of $\mathbb{P}_{\rm fail}(t; W,L)$ is simply $1- 2^{-(\# \text{logicals})}$,
which for the repetition code equals $1/2$.}.
We thus define $t_{\rm mem} \equiv \Gamma_{W,L}^{-1}$.
In \cref{fig:placeholder_fig_1}(c), we numerically find that 
\begin{align}
    \label{eq:Gamma_finite_size_scaling}
    - L^{-1} \cdot \ln (\Gamma_{W,L} / \Gamma_0) \propto \Phi(W/L),
\end{align}
in agreement with \cref{eq:crossover_scaling}.
The limiting behaviors of the functional form at small and large values of $W/L$ also agree with our expectations below \cref{eq:crossover_scaling}.
The growth rate of $\Delta \rho(t; W, L)$ obeys a similar scaling collapse.

In performing the data collapse, the only fitting parameter is the overall multiplicative factor $\Gamma_0$ of $\Gamma_{W, L}$ contributing to an \textit{additive} constant to the RHS of~\cref{eq:crossover_scaling}, which we fix independently using data from the static limit ($W\rightarrow\infty$). We expect the $p$-dependence of the failure probability to be captured entirely by the ``line tension'' $L^{-1} \cdot \ln (\Gamma_{W,L}/\Gamma_0)$, and $\Gamma_0$ to be a geometric factor independent of $p$. 
Therefore, for all three values of $p$ in \cref{fig:placeholder_fig_1}(c), we use the same fitting parameter $\Gamma_0$.

We note that the form in \cref{eq:crossover_scaling} also allows for a non-universal \textit{multiplicative} constant between the LHS and the RHS.
Indeed, data at different values of $p$ can be collapsed onto the same function $\Phi$, after an overall rescaling of the vertical axis, confirming our expectation that $\Phi$ is universal~\footnote{We note that a rescaling of the horizontal axis is not required to collapse data at different $p$ in~\cref{fig:placeholder_fig_1}, as long as the decoder is fixed.
Comparing the asymptotics of the scaling function $\Phi(x)$ [see discussions below~\cref{eq:crossover_scaling}], namely
$\Phi(x \to \infty) = \sigma(p)$ and $\Phi'(x=0) = \alpha \kappa \sigma(p)$, we see that the dependence on $p$ comes entirely from the line tension, and can thus be eliminated by rescaling the vertical axis only.
In particular, the ``bend'' of $\Phi$ is near $x_\ast \approx 1/(\alpha \kappa)$, independent of $p$.}.
Closer to the critical point $p = p_c$, the vanishing of the line tension should be universal, but
the presence of multiple length scales makes a numerical calculation more demanding, and we leave a more thorough investigation of the critical regime for future work.

\begin{figure}[t]
    \centering
    \includegraphics[width=\linewidth]{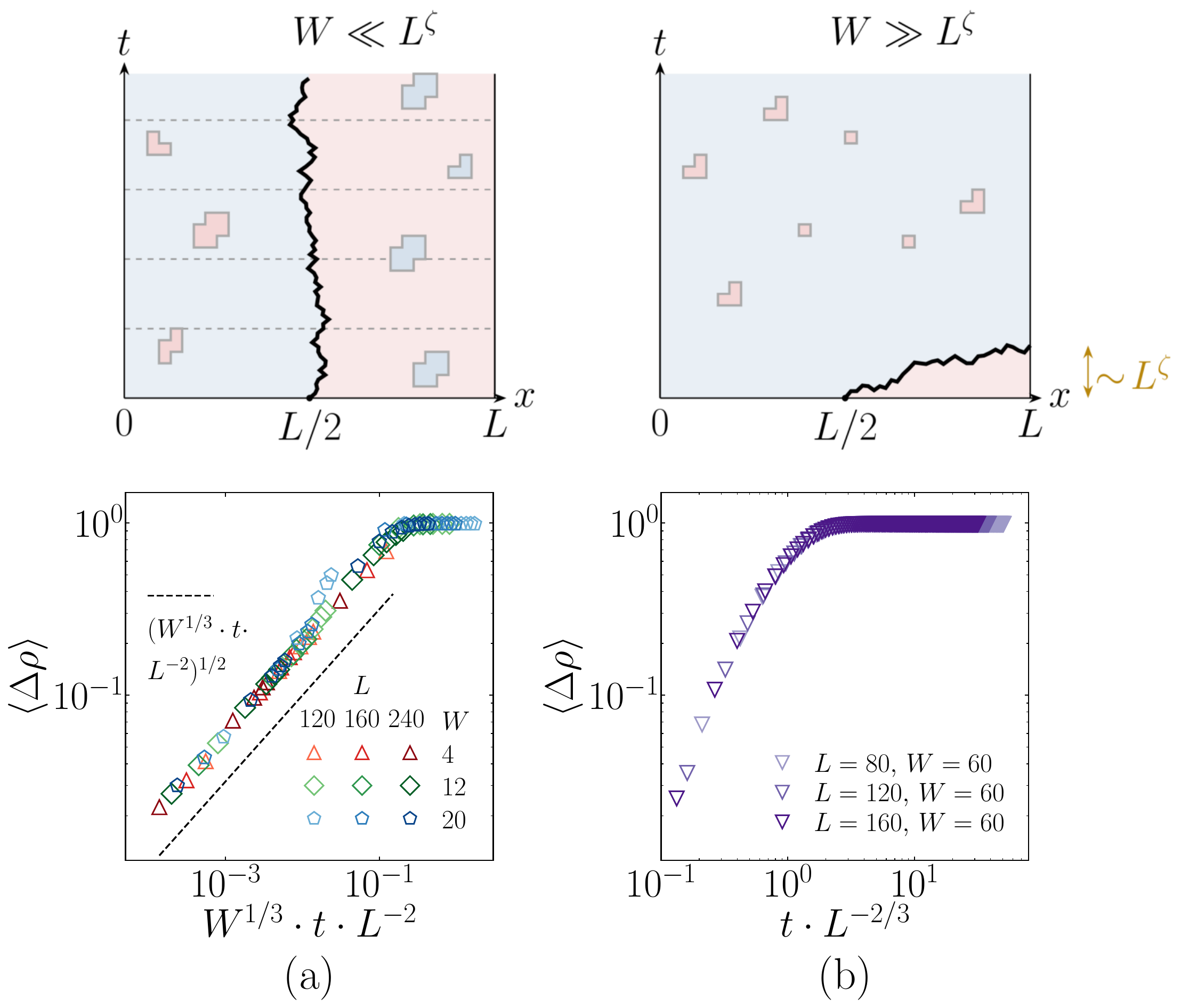}
    \caption{Diffusive motion of $\mathbb{Z}_2$ charges.
    (a) For $W \ll L^\zeta$, we predict that $\avg{|\Delta \rho|} \propto (D \cdot t \cdot L^{-2} )^{1/2}$, where $D \propto W^{1/3}$ is the effective diffusion constant, see \cref{eq:diffusion_constant_result}.
    (b) For $W \gg L^\zeta$, the domain wall tends to take a shortest path to terminate either on the left or right boundary.
    Thus, $\avg{\Delta \rho}$ saturates at time $t \propto L^\zeta$, where $L^\zeta$ is the height of the domain wall.
    At earlier times, $\avg{\Delta \rho}$ depends on $t$ through the scaling variable $t \cdot L^{-\zeta}$.
    }
    \label{fig:placeholder_fig_2}
\end{figure}

Next, we establish numerical evidence for the diffusive scaling of the dynamics and extract the diffusion constant, see \cref{eq:diffusion_constant_result}.
We initialize the system with a single domain wall at $x = L/2$ (so that $\Delta \rho = 0$), as illustrated in \cref{fig:placeholder_fig_2}, and study its motion under SWD. 
For this experiment we take open boundary conditions.

We first focus on the regime $W \ll L^\zeta$ [\cref{fig:placeholder_fig_2}(a)]. Here, from discussions leading to \cref{eq:diffusion_constant_result},
the isolated domain wall appears as a random walker, with directed polymers of temporal extent $\propto W$ as its microscopic steps.
The distance of the random walker from its initial location $x=L/2$ can be related to the spin density as 
\begin{align}
    \langle |\Delta x| \rangle
    \propto
    L \cdot \langle \Delta \rho 
    \rangle
    ,
\end{align}
where $\langle \cdot \rangle$ is the expectation over 
all possible error configurations weighted by the phenomenological noise model.
This observable has the advantage that fast pair creation-annihilation processes (i.e. bubbles whose size distribution is independent of $W$) do not contribute. For diffusive scaling $\avg{\abs{\Delta x}} \propto (D(W) \cdot \Delta t)^{1/z}$, we expect that
\begin{align}
    \avg{\Delta \rho}
    \;
    \propto
    \;
    L^{-1} \avg{\abs{\Delta x}}
    \;
    \propto 
    \;
    L^{-1} (W^{z\zeta - 1} \cdot \Delta t)^{1/z},
\end{align}
where we have used $D \propto W^{z\zeta - 1}$ \eqref{eq:diffusion_constant_result}.
Our data is in excellent agreement with this expectation (see \cref{fig:placeholder_fig_2}(a)).

In the opposite regime $W \gg L^\zeta$ (see \cref{fig:placeholder_fig_2}(b)), the dominant contributions come from DPRMs directed in the spatial direction. Its temporal extent scales as $L^\zeta$~\cite{Note2}
and for $t\gg L^\zeta$ the random walker has a high probability of having terminated on either the left or right boundaries, with $\avg{\Delta \rho}$ saturating at its steady state value. 
Before saturation, the growth of $\avg{\Delta \rho}$ is described by a scaling function with a single scaling variable $t \cdot L^{-\zeta}$.

These two experiments both support the theoretical picture we outlined, including the reaction-diffusion \cref{eq:RD} as well as the scaling of the pair nucleation rate~\cref{eq:tau_W_scaling}, the diffusion constant~\cref{eq:diffusion_constant_result}, and the memory time~\cref{eq:crossover_scaling}.

\emph{Discussion} --- %
In this work, we have studied the dynamics of SWD, which is one type of real-time QEC, focusing on topological codes with deconfined point-like defects.
We have argued that the long-time, large-scale behavior admits a simple effective description, namely a parity-conserving reaction-diffusion process.
The memory time of the code is the relaxation time of this effective process, and its dependence on $W$ and $L$ is captured by a single-variable crossover scaling that interpolates between the reaction-diffusion regime at $W \ll L$ and the static regime at $W \gg L$.
We argued for this general description, and provided extensive numerical evidence for its validity, using the 1D repetition code and the 2D toric code (End Matter) as numerical examples.

Our results provide concrete theoretical guarantees for the design and evaluation of SWD-style real-time decoders.
First, the crossover scaling identifies $W/L$ as the natural figure of merit for speed-accuracy tradeoff, which can be used to inform error-budgeting with a given throughput constraint.
Second, the reaction-diffusion picture is general and applies to a broad range of noise models and decoders, %
including those with explicit quasi-locality in both space and time or those with lower algorithmic complexity.
We explore these possibilities in the End Matter and Supplemental Material~\cite{Note2}, finding agreements with the reaction-diffusion process.

The dynamics of SWD is sensitive to the sizes of the commit and buffer windows, which control the creation of the slow $\mathbb Z_2$ charges whose subsequent diffusion across the system leads to logical failure.
Previous numerical studies of SWD~\cite{skoric2023parallel,tan2022parallel,bombin2023modular} found that logical error suppression is exponential in the code distance $d$ and performance close to the static setting can be recovered for windows scaling linearly with $d$.
Our analysis of SWD provides a dynamical explanation for this dependence;
varying the window sizes allows to probe sensitivity to timelike measurement errors, closely connecting to stability experiments~\cite{Gidney2022} and lattice-surgery simulations~\cite{Chamberland2022}.

Our effective model provides a new perspective on the fact that SWD is not a single-shot decoder for codes with point-like excitations.
For any finite $W$, the memory time saturates and does not diverge as a function of $L$.
Indeed, from~\cref{eq:crossover_scaling}, the rate of decoding $1/W$ acts as a \textit{relevant perturbation} to the decodable (ordered) phase, and similarities can be drawn with dimensional crossover~\cite{Cardy_book}, see also~\cite{tan2025burst, adithya2024rareevents}.
Single-shot decoders, on the other hand, have diverging memory times even when the decoder only has access to syndrome information from a single time step~\cite{bombin2015singleshot, KubicaVasmer, BrownNickersonBrowne2016GCC, YL2023STC, stahl2023singleshot}.
Even though allowing for $W=O(1)$ time steps may improve the threshold~\cite{huang2023increasing}, it does not qualitatively change the behavior of the decoder.
An effective description of single-shot decoding would provide a universal understanding of this behavior. In particular, it would allow for the more general definition of single-shot memories as those for which $1/W$ is an irrelevant perturbation to the memory, unlike SWD where $1/W$ is relevant.

Looking forward, real-time QEC may be viewed as perhaps the simplest non-trivial example of many-body dynamics driven by local operations and classical communication (LOCC).
The resulting dynamics is neither unitary nor described by a fixed local quantum channel.
What other kinds of many-body dynamics can be realized under LOCC, when the figure of merit are communication bandwidth, range, latency, in addition to the relaxation time? 
Which non-equilibrium universality classes, if any, are accessible to LOCC but not to local rules?
While examples in this work belong to a known dynamical universality class,
excitations carrying charge in a more general group than $\mathbb{Z}_2$ (either abelian or non-abelian), or excitations that obey kinetic constraints (instead of being deconfined)~\cite{PhysRevE.60.5068, PhysRevA.83.042330}, could lead to qualitatively new behavior.

\emph{Acknowledgments} --- %
We thank Arpit Dua, Hyunsoo Ha, Jacob Hauser, Vedika Khemani, Ethan Lake, Aditya Mahadevan, Nicholas O'Dea, and Shengqi Sang for helpful discussions.
A.S. was supported in part by the Office of Naval Research Young Investigator Program (ONR YIP) under Award Number N00014-24-1-2098 and in part by a Packard Fellowship in Science and Engineering (PI: Vedika Khemani). A.S. also acknowledges support from the NSF graduate research fellowship and the ARCS Scholar award.
C.S. was supported by the US Department of Energy, Office of Science under Award No DE-SC0025934.
A.K. acknowledges support from the NSF (QLCI, Award No. OMA-2120757), IARPA and the
Army Research Office (ELQ Program, Cooperative Agreement No. W911NF-23-2-0219).
Y.L. was supported in part by the Gordon and Betty Moore Foundation's EPiQS Initiative through
Grant GBMF8686, in part by the US Department of Energy, Office of Science under Award No DE-SC0019380, and in part by a Q-FARM Bloch Postdoctoral Fellowship at Stanford University.
We acknowledge the hospitality of the Kavli Institute for Theoretical Physics (KITP), which is supported in part by grant NSF PHY-2309135.

\let\oldaddcontentsline\addcontentsline%
\renewcommand{\addcontentsline}[3]{}%

\bibliographystyle{apsrev4-2}
\bibliography{refs}

\clearpage

\section*{End matter }\label{sec:end_matter}

Here, we explore a number of different scenarios for SWD beyond the repetition code and MWPM, with the aim of extending the validity of our effective description to a greater range of codes and decoders.

\subsection{Explicitly quasi-local dynamics}

\begin{figure}[b]
    \centering
    \includegraphics[width=\linewidth]{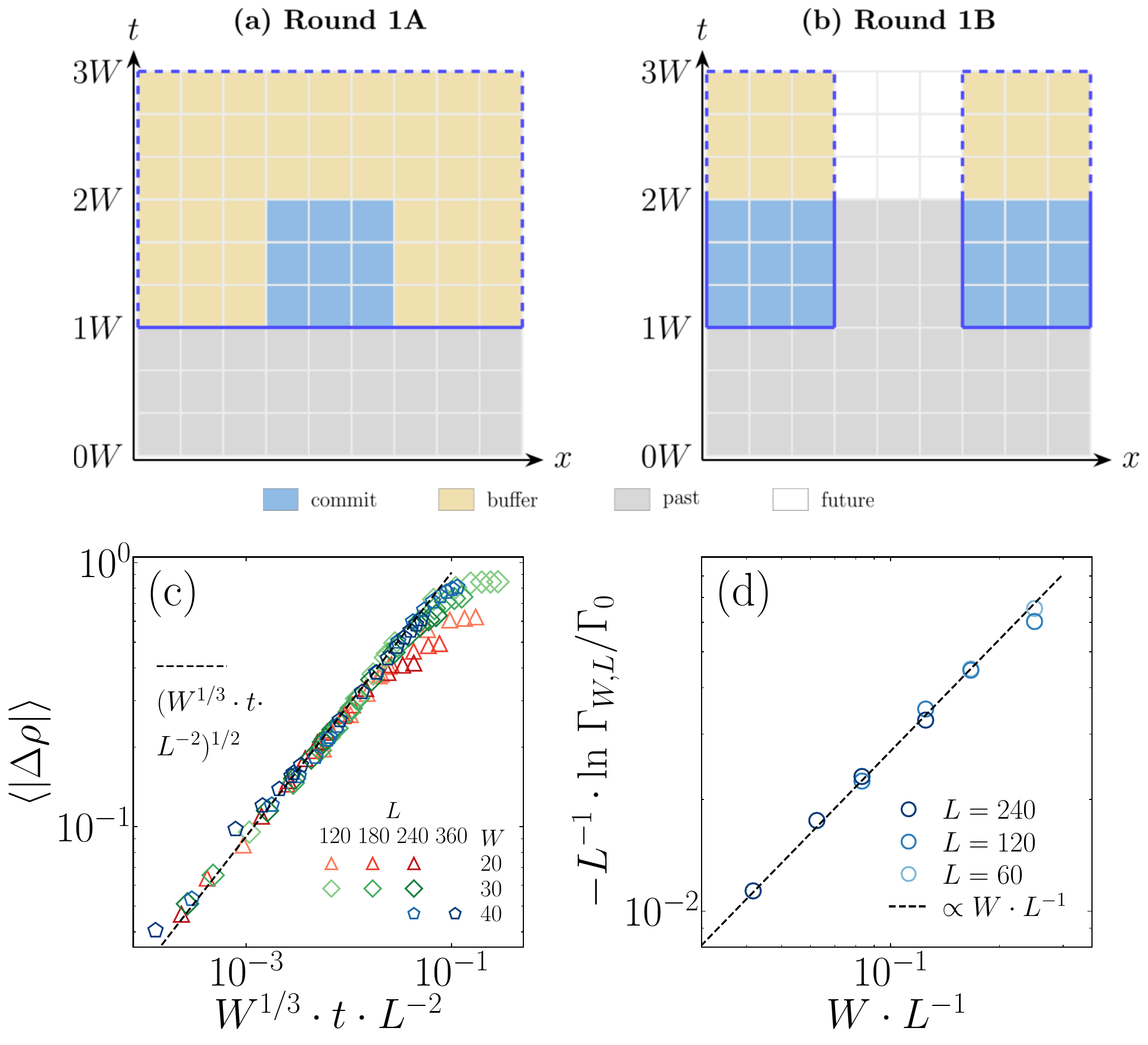}
    \caption{(a, b) Layout of modular SWD. (c) Results for diffusive motion under modular SWD, compare with~\cref{fig:placeholder_fig_2}(a).
    (d) Results for the crossover scaling of the memory time under modular SWD, compare~\cref{fig:placeholder_fig_1}(c).}
    \label{fig:SBD}
\end{figure}

The quasi-local effective description, particularly that point defects have a ``range of interaction'' of $O(W)$, suggests that it might be possible to capture large scale properties of SWD (which is nonlocal in space) by a microscopic model with quasi-locality of range $O(W)$.
We examine a few such realizations and confirm these expectations.
These numerical experiments serve as additional nontrivial tests of our general reaction-diffusion description.
They also provide theoretical guarantees to these decoders (which  are desirable from the practical standpoint due to their quasi-locality and numerical efficiency).
In all cases, we focus on $\mathbb{P}_{\rm fail}(t; W, L)$ as described in the main text.

\begin{figure}[b]
    \includegraphics[width=\linewidth]{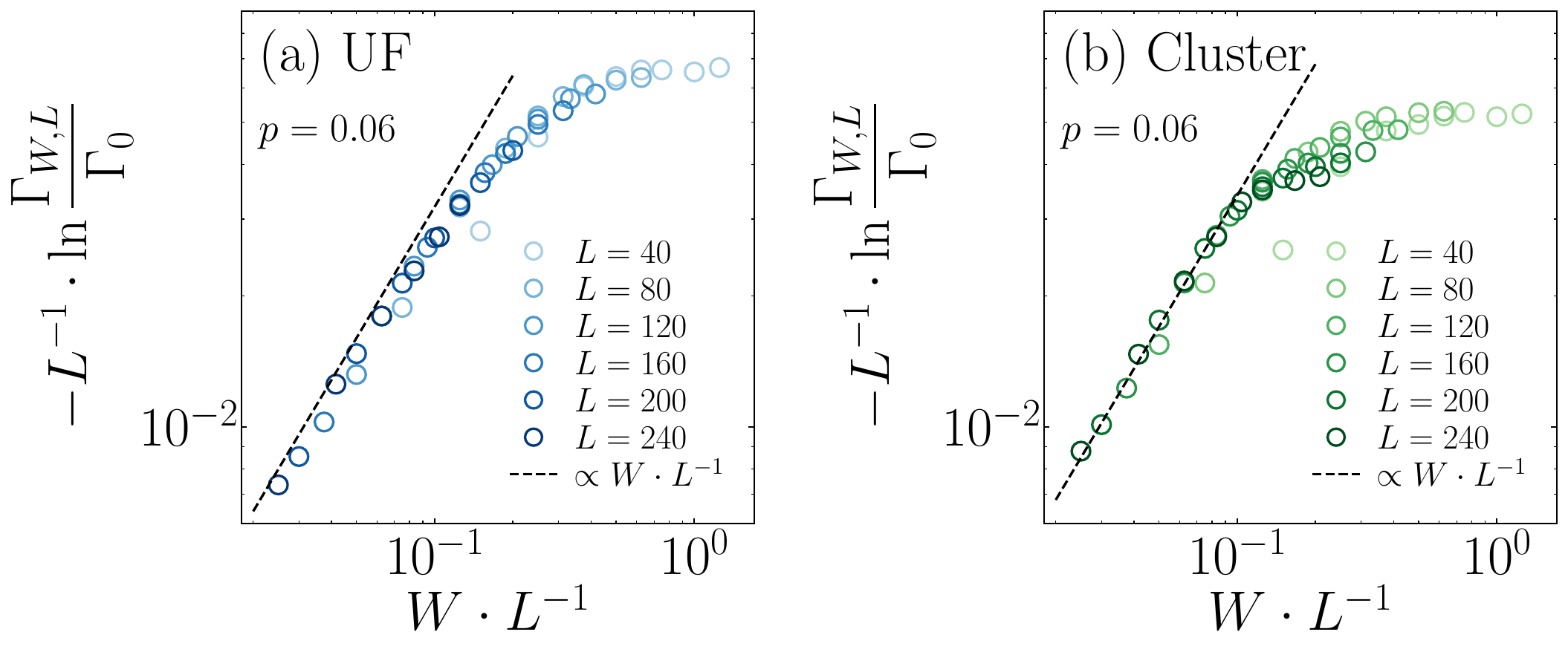}

    \caption{Results from $\mathbb{P}_{\rm fail}$ of SWD with the UF and clustering decoders.
    The data is in agreement with~\cref{eq:crossover_scaling} and~\cref{fig:placeholder_fig_1}(c).
    }
    \label{fig:uf_decoder}
\end{figure}

\subsubsection{Modular SWD \label{sec:SBD}}

We first consider the modular SWD, defined in \cref{fig:SBD}(a,b).
We divide the spacetime lattice into square blocks of size $W \times W$.
Within each round, we first commit to even-numbered blocks, with their $3W \times 2W$ surrounding as a buffer (\cref{fig:SBD}(a)); and then commit to odd-numbered blocks, with their $W \times 2W$ surrounding as a buffer (\cref{fig:SBD}(b)).
Because the even blocks have already been committed, the odd blocks require a smaller buffer.
Attention must be paid to the boundaries of the decoding windows: dashed lines between \textit{buffer} and \textit{future} indicates that the correction chain $E'$ is allowed to cross that boundary, whereas solid lines between \textit{commit} and \textit{past} indicates that the correction chain $E'$ is constrained not to cross that boundary.
Similarities can be found between modular SWD and the modular decoder proposed in Ref.~\cite{bombin2023modular}.

Our results are shown in \cref{fig:SBD}(c,d).
In \cref{fig:SBD}(c), we find similar scaling of the diffusion constant $D$
as in \cref{eq:diffusion_constant_result} and \cref{fig:placeholder_fig_2}(a).
Due to the lower computational cost, we can access larger system sizes while maintaining $W \ll L^\zeta$, so that the diffusive scaling can be verified for a larger range of $W$.
In \cref{fig:SBD}(d), we find similar crossover scaling of $t_{\rm mem}$ as in \cref{eq:crossover_scaling} and \cref{fig:placeholder_fig_1}(c).
By design, we are limited to regimes $W / L \ll 1$, as the blocks have spatial extent limited by the system size; this explains why we do not see the full range of the scaling function~$\Phi$. 

We thus conclude that modular SWD correctly captures universal aspects of SWD, while requiring a smaller numerical overhead, as it operates on an $O(W) \times O(W)$ decoding window, rather than an $O(L) \times O(W)$ one.
Due to its modularity,
modular SWD should be the preferred choice over SWD for real-time decoding.

\subsubsection{Union-Find and Clustering decoders}

Next, we consider the 1D repetition code and SWD, but with MWPM replaced by either the union-find (UF)~\cite{Delfosse2021almostlineartime} or clustering~\cite{poulin2010RGdecoder,bravyihaah2013} decoders.
Both algorithms proceed by growing clusters isotropically from seeds of point-like defects.
Clusters stop growing whenever they become resolvable by an error chain, which in our convention are allowed to terminate on the boundary of a window.
Comparing the two, the clustering decoder removes a resolvable cluster by finding and applying an error chain
immediately after one is found, whereas the UF decoder performs global removal at the very end.

Under the UF or clustering decoders, pairs of $\mathbb{Z}_2$ charges separated by a distance greater than $O(W)$ are essentially invisible to each other.
Therefore, similarly to SWD with MWPM, SWD with the UF or clustering decoders is also expected to undergo quasi-local motion as described by \cref{eq:RD}.
Indeed, with numerical results in~\cref{fig:uf_decoder} we find agreements with the crossover scaling in~\cref{eq:crossover_scaling} and~\cref{fig:placeholder_fig_1}(c).
Note that the error rate is adjusted accordingly, as the UF and clustering decoders have a lower threshold.

\subsection{Crossover scaling for the 2D toric code \label{sec:2DTC}}

\begin{figure}[b]
    \centering
    \includegraphics[width=\linewidth]{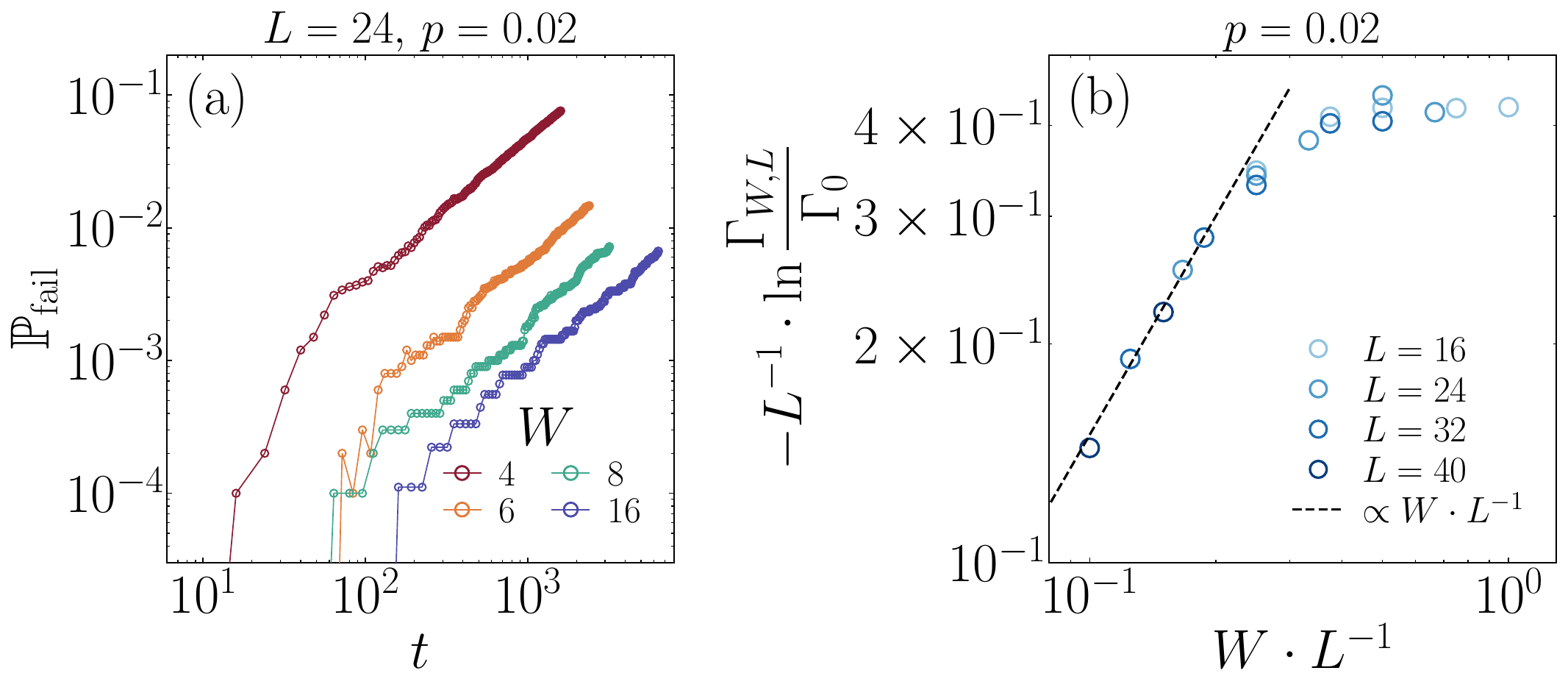}
    \caption{Crossover scaling of the 2D toric code (compare with \cref{fig:placeholder_fig_1} for the 1D repetition code).
    For early times, we did not observe any logical failures in $10^{4}$ realizations. 
    }
    \label{fig:2DTC}
\end{figure}

We now examine whether the reaction-diffusion description established for the 1D repetition code in the main text extends to the 2D toric code.
The code is defined on an $L \times L$ square lattice with periodic boundary conditions, and is subject to phenomenological noise: independent bit-flip (Pauli Z) errors on data qubits at rate $p$ and independent measurement errors on syndrome bits at rate $p$.
We decode using SWD with MWPM acting within each window, following the same commit/buffer/projection prescription as in \cref{fig:placeholder_fig_1}(a).

The microscopic degrees of freedom differ from those of the 1D repetition code in a few ways.
On each time slice, the syndrome is a collection of violated plaquette operators, i.e. point-like $\mathbb{Z}_2$ anyons.
Therefore, $E \oplus E'$ is a collection of closed loops in $(2{+}1)$-dimensional spacetime.
(A similar picture holds for the phase-flip (Pauli X) errors.)
The coarse-graining arguments nevertheless go through: syndrome pairs separated by more than $W$ in spacetime are not resolved within a single decoding window, so the screening cutoff is still of order $W$, and the slow degrees of freedom of the effective theory are coarse-grained anyons undergoing parity-conserving reaction-diffusion process.

The main differences from the 1D case are that the anyons now diffuse in two spatial dimensions.
Therefore, the scaling of the diffusion constant $D(W) \propto W^{z \zeta - 1}$ will have a different value of $\zeta$, given by a directed polymer in three spatial dimensions.
However, we have not been able to directly test this.
The (1+1)D domain-wall order parameter $\Delta \rho$
used to track a coarse-grained charge in \cref{fig:placeholder_fig_2} has no direct (2+1)D analogue, and the position of an isolated coarse-grained charge cannot be read off from a local observable in the same way.
For this reason, we do not present a test of the diffusion scaling for the
toric code analogous to \cref{fig:placeholder_fig_2}, and instead focus on the crossover scaling of the memory time.

Initializing the system in a logical codeword state we run SWD %
at fixed error rate $p < p_c$ for long enough to extract the logical failure rate $\Gamma_{W,L}$ from the linear growth of $\mathbb{P}_{\rm fail}(t; W, L)$, as in the main text.
We find that the same scaling form applies,
\begin{align}
    - L^{-1} \cdot \ln \frac{\Gamma_{W,L}}{\Gamma_0} = \Phi_{\rm 2D}(W/L),
\end{align}
where $\Phi_{\rm 2D}$ has the same asymptotic behavior as the 1D scaling function in \cref{eq:crossover_scaling}, i.e. linear at small argument and saturating at large argument, see \cref{fig:2DTC}(b).
We note that the static threshold for the 2D toric code under phenomenological noise is $p_c \approx 2.9\%$~\cite{DKLP2001topologicalQmemory, WangHarringtonPreskill2003}, lower than that of the 1D repetition code at the same abstract level of description; this limits the range of error rates over which the scaling can be tested deep in the decodable phase.

\let\addcontentsline\oldaddcontentsline%

\cleardoublepage
\newpage

\onecolumngrid
\begin{center}
\textbf{\large Supplemental Material for ``Kinetics of sliding-window quantum error correction''}
\vskip 0.4cm
{Adithya{\;\,}Sriram,\textsuperscript{1}{\;\,}Charles{\;\,}Stahl,\textsuperscript{1}{\;\,}Aleksander{\;\,}Kubica,\textsuperscript{2}{\;\,}and{\;\,}Yaodong{\;\,}Li\textsuperscript{1,\,3}}
\vskip 0.1cm
\textsuperscript{1}{\fontsize{9.5pt}{11.5pt}\selectfont\emph{Department of Physics, Stanford University, Stanford, CA 94305}}\\[-0.75pt]
\textsuperscript{2}{\fontsize{9.5pt}{11.5pt}\selectfont\emph{Yale Quantum Institute \& Department of Applied Physics, Yale University, New Haven, CT 06520}}\\[-0.75pt]
\textsuperscript{3}{\fontsize{9.5pt}{11.5pt}\selectfont\emph{Department of Mechanical Engineering \& Department of Physics, University of Washington, Seattle, WA 98195}}\\[-0.75pt]
{\fontsize{9pt}{11pt}\selectfont(Dated: August 11, 2026)}
\vskip 0.75cm
\end{center}

\setcounter{secnumdepth}{2}

\setcounter{equation}{0}
\setcounter{figure}{0}
\setcounter{table}{0}
\setcounter{page}{1}
\makeatletter
\renewcommand{\thesection}{S\arabic{section}}
\renewcommand{\theequation}{S\arabic{equation}}
\renewcommand{\thefigure}{S\arabic{figure}}

\section{Formal definition and effective description of the sliding window decoder (SWD)}

In this section we describe the SWD in some more detail, and provide motivations for the effective description in \cref{eq:RD}.

\subsection{Formal definition of the SWD}

Our discussion is focused on $d$-dimensional topological CSS codes with deconfined point-like defects; see Ref.~\cite{DKLP2001topologicalQmemory} for more details.
In this setting, the decoding problem can be associated with a $(d+1)$-dimensional spacetime lattice.
The spacetime locations of physical errors and measurement errors are captured by the \textit{error chain}; the \textit{correction chain} captures the output of the local decoder $\mathcal{R}$, i.e., the locations of estimated errors.
The detectors, which, in the absence of errors, are deterministic products of stabilizers, can be associated with the vertices of the spacetime lattice.
The syndrome information corresponds to the triggered detectors (note that a detector is triggered only if the number of errors incident to it is odd).
For the error chain $E$, we denote the corresponding syndrome as $\partial E$.
For the specific cases of the repetition code and the $d$-dimensional toric code (with point-like defects), each individual error triggers at most two detectors, and therefore we can talk about $\mathbb{Z}_2$ charges at the endpoints of the error chain.

The task of the decoder is to output a correction consistent with the observed syndrome (or, in other words, a correction that annihilates the defects), while trying to minimize the probability of causing a logical error.
This correction $E' = \mathcal{R}(\partial E)$ will naturally have common endpoints with $E$, namely $\partial E = \partial E'$; these endpoints $\partial E$ correspond to ``detection cells'' (also known as frustrated plaquettes in an associated stat-mech model), which serve as inputs to $\mathcal{R}$.
Their combination $E \oplus E'$ forms closed loops on the spacetime lattice, which for (1+1)D can be interpreted as domain walls of an effective Ising spin configuration.
The closed loops in $E \oplus E'$ can also be viewed as worldlines of $\mathbb{Z}_2$ charges.
Generally speaking, while the error chain $E$ can be defined unambiguously and independently of any decoder, the correction chain $E'$ must be defined with respect to a decoder.

The sliding window decoder (SWD) operates in rounds, labeled by a non-negative integer $n \in [0, N]$.
Define the cumulative correction chain before round $n$ to be
\begin{align}
    (E'_\CC)^{<n}  \coloneqq \bigoplus_{n' < n} E'_\CC(n').
\end{align}
Here, $E'_\CC(n)$ is the \textit{committed} correction chain in round $n$, to be defined inductively below.
By convention, $(E'_\CC)^{(<0)} = \emptyset$ and $E'_\CC(n) = \emptyset$ for all $n < 0$.

During the $n$-th round, the decoder inputs syndromes within the thin strip in spacetime (see \cref{fig:placeholder_fig_1}).
\begin{align}
    B(n) \coloneqq [n W, (n+2) W] \times [0, L]
\end{align}
Within $B(n)$, the decoder performs decoding based on the observed 
detection cells as follows
\begin{align}
    \label{eq:Eprime_tot_def}
    E'(n) \coloneqq \mathcal{R} \ld \partial \lz \( E \oplus (E'_\CC)^{<n} \) \cap B(n) \rz \rd.
\end{align}
Notice that the syndrome is a result of both physical error $E$ and past corrections $(E'_\CC)^{<n}$.
Define the \textit{commit window}
\begin{align}
    B_\CC(n) \coloneqq [n W, (n+1) W] \times [0, L]
\end{align}
and the \textit{buffer window},
\begin{align}
    B_\BB(n) \coloneqq [(n+1) W, (n+2) W] \times [0, L].
\end{align}
The decoder only commits to corrections in $B_\CC(n)$,
\begin{align}
    \label{eq:Eprime_commit_def}
    E'_\CC(n) \coloneq E'(n) \cap B_\CC(n),
\end{align}
whereas $B_\BB(n)$ serves as a look-ahead that improves decoding accuracy, but no final decisions are made there yet.
As the commitment involves a change of basis on the physical space, the syndrome data will need to be transformed covariantly.

After committing to $E'_\CC(n)$, the decoder discards the syndrome data from $B_\CC(n)$, and $B_\BB(n)$ slides forward to become $B_\CC(n+1)$.
Each round thus overlaps with its neighboring ones, and corrections are made every $W$ time steps at a rate $R \propto 1/W$.

Special attention to boundary condition is necessary when 
performing $\mathcal{R}$ within $B(n)$.
The correction chain $E'(n)$ might span both $B_\CC(n)$ and $B_\BB(n) = B_\CC(n+1)$, but the decoder will need to erase everything within $B_\BB(n)$ in the next round.
Therefore, we must allow the possibility of \textit{projecting} detection cells in $B_\CC(n)$ onto the temporal boundary between $B_\CC(n)$ and $B_\BB(n) = B_\CC(n+1)$, see \cref{eq:Eprime_commit_def}, where $E'_\CC(n)$ effectively implements this projection. In this way, the projected detection cells can be picked up in the next round, see \cref{eq:Eprime_tot_def}, where $\mathcal{R}$ receives input from $(E'_\CC)^{<n}$, including any such projection. This ensures that $E \oplus E'$ form a closed loop, from which a consistent error history can be reconstructed.

\subsection{Quasi-local motion of coarse-grained $\mathbb{Z}_2$ charges}

Viewing the closed loops in $E \oplus E'$ as worldlines of $\mathbb{Z}_2$-charged point defects, we now argue that their motion admits a quasi-local description, as long as we look at length scales and time scales long compared to $W$.

A quasi-local description is not obvious \textit{a priori}, as the decoder $\mathcal{R}$ performs a global matching of point-like detection cells, and $E'(n) \coloneqq \mathcal{R}[\partial E \cap B(n)]$ can in principle be long-ranged.
One might therefore expect that any effective description of the evolution of $E \oplus E'$ will necessarily be nonlocal as well.

However, because the detection cells are allowed to be paired with temporal boundary (see \cref{fig:placeholder_fig_1}(a)), the window width $W$ introduces a cutoff on the length of $E'(n)$.
Indeed, consider a defect pair separated by $\Delta x \gg W$ in space.
Within a single window, $\mathcal{R}$ can either match the two endpoints with each other (at cost $O(\Delta x)$) or match each endpoint locally with the temporal boundary (at cost $O(W)$).
When $\Delta x \gg W$, the latter is always favorable, and the two endpoints are effectively ``screened''.
More generally, any connected component of length $\gg W$ in $E'(n)$ will be suppressed exponentially.

Therefore, we have argued that $E'(n) \coloneqq \mathcal{R}[\partial E \cap B(n)]$ will typically have connected components of lengths at most $O(W)$.
Therefore, $E_\CC'(n) \subseteq E'(n)$ also typically has connected components of lengths at most $O(W)$.
Writing 
\begin{align}
    \label{eq:gamma_def}
\gamma_{\rm tot} \coloneqq E \oplus E'=
\bigoplus_n  \underbrace{\lz E \cap B_\CC(n) \rz  \oplus E'_\CC(n)}_{\eqcolon \gamma(n)}
\end{align}
and interpreting them as worldlines of $\mathbb{Z}_2$ charges, our argument implies that the $\mathbb{Z}_2$ charges also move in a quasi-local manner.
In particular, the time evolution within the round $n$  is determined by $\gamma(n)$, which contains local moves of size at most $O(W)$.

Nevertheless, the charges can still travel a long distance, if they survive over many rounds, see~\cref{fig:placeholder_fig_1}(b).

\section{Branching-annihilating random walkers (BARW) under SWAP noise channel \label{sec:BARW-SWAP}}

The reaction-diffusion description developed in the main text does not have an absorbing-state phase transition, because the noise in the standard SWD setting is always present and the empty configuration is never a strict fixed point of the dynamics.
Here we describe a modified error model in which the absorbing state is indeed stable, allowing access to an absorbing state phase transition.
This provides another check of our effective description.

We consider the 1D repetition code under a noise model in which
bit-flip errors are turned off, and the only physical noise is
a random SWAP channel acting on neighboring bits.
A SWAP acts trivially on  uniform configurations of all $0$ or all $1$, therefore preserve the absorbing states.
In the language of \cref{eq:RD}, pair creation is forbidden
($\tau = 0$).
When there is an isolated domain wall, a SWAP acting across the domain wall can move a $0$ into the region of $1$s (or vice versa), creating an isolated bit surrounded by its opposite and hence two new domain walls,
\begin{align}
    00001111 \to 00010111.
\end{align}
Under the joint action of the SWAP channel (which supplies the 1-to-3 branching of domain walls) and the SWD feedback (which matches syndromes and contributes to annihilation on contact, as well as to diffusion), the domain walls evolve as a parity-conserving  branching-annihilating random walk (BARW)~\cite{CardyTauber1998, hinrichsen2000DP},
\begin{subequations}
\label{eq:BARW}
\begin{align}
   \emptyset A &\xleftrightarrow{D} A \emptyset,
   \\
   AA &\xrightarrow{\lambda} \emptyset,
   \\
   A &\xrightarrow{\sigma} AAA.
\end{align}
\end{subequations}
which is known to undergo an absorbing-state phase transition
in the DP2 (parity-conserving) universality class as $\sigma$
is tuned.
Note that in the present setting $D$ cannot be varied independently of $\sigma$.

\begin{figure}[t]
    \centering
    \includegraphics[width=0.95\linewidth]{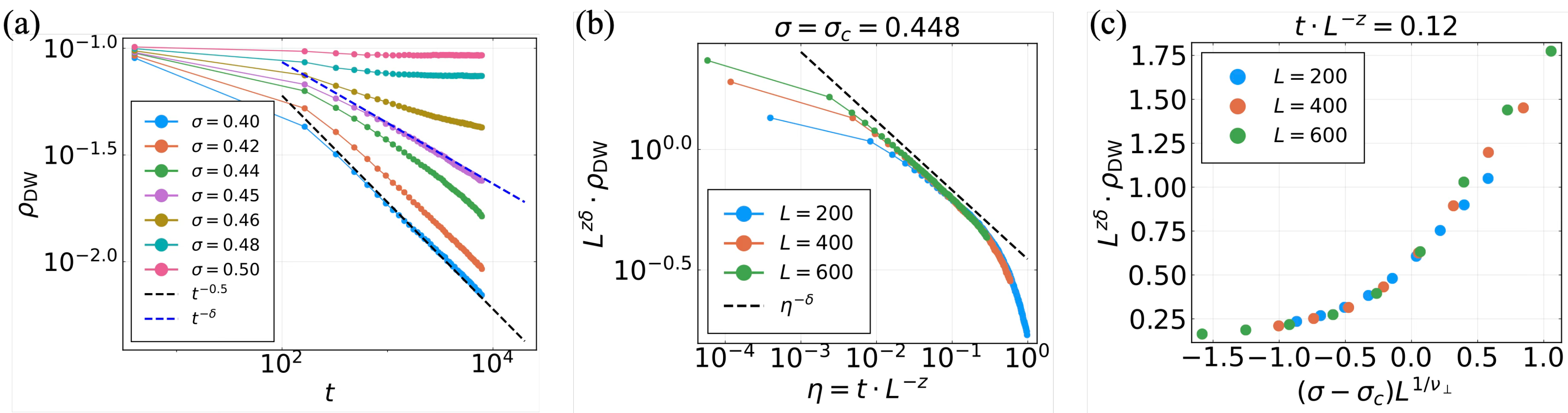}
    \caption{Numerical results for the SWD under a SWAP noise channel, which we compare with a branching-annihilating process, see~\cref{eq:BARW}.
    (a) When we increase the SWAP rate $\sigma$, the system goes from an \textit{absorbing} phase where the domain wall density decays as $\rho_{\rm DW} \propto t^{-1/2}$ at long times, to an \textit{active} phase where $\rho_{\rm DW}$ saturates at a nonzero constant.
    (b, c) The critical data for $\rho_{\rm DW}$ can be fitted to~\cref{eq:BARW_scaling_function}, and agreements with DP2 exponents are found.
    }
    \label{fig:BARW}
\end{figure}

We simulate this model and tune through an absorbing-state transition.
We fix $W = 4$, take the SWD with MWPM as its inner decoder, and start with an initial state with domain wall density $\rho_{\rm DW} = 0.2$.
For $\sigma < \sigma_c$, the system is in the absorbing phase and we find the diffusion-limited asymptotic decay $\rho_{\rm DW} \propto t^{-1/2}$, see Fig.~\ref{fig:BARW}(a).
Near the critical point $\rho_{\rm DW}$ obeys the following standard scaling relation,
\begin{align}
    \label{eq:BARW_scaling_function}
    \rho_{\rm DW}(\sigma; t, L) = t^{-\delta}\, \Upsilon(t\cdot L^{-z}, (\sigma-\sigma_c) \cdot L^{1/\nu_\perp}) = 
    L^{-z\delta}\, \widetilde{\Upsilon} (t\cdot L^{-z}, (\sigma-\sigma_c) \cdot L^{1/\nu_\perp}),
\end{align}
where $\delta \approx 0.285, \nu_\perp \approx 1.83, z \approx 1.74$ are exponents of the DP2 universality class~\cite{ZHONG1995333}.
The scaling ansatz as well as the DP2 exponents are consistent with our findings in Fig.~\ref{fig:BARW}(b,c).
We view this as a further nontrivial test of our effective description, where the reaction-diffusion process is tuned across a critical point.

Our results should be generally applicable for any $W = O(1)$.
The crossover to large $W$ is an interesting question we leave for the future.

\end{document}